\documentclass[fleqn,usenatbib]{mnras}

\usepackage{newtxtext,newtxmath}

\usepackage[T1]{fontenc}

\DeclareRobustCommand{\VAN}[3]{#2}
\let\VANthebibliography\thebibliography
\def\thebibliography{\DeclareRobustCommand{\VAN}[3]{##3}\VANthebibliography}

\usepackage{graphicx}	
\usepackage{amsmath}	

\title[21-cm PS sensitivity to DM–baryon scattering]{Forecasting 21-cm power spectrum sensitivity to dark Matter-baryon scattering}

\author[A. Rahimieh, P. Parashari and V. Gluscevic]{
Aryan Rahimieh,$^{1}$\thanks{E-mail: rahimieh@usc.edu}
Priyank Parashari$^{1}$\thanks{E-mail: ppriyank@usc.edu}
and Vera Gluscevic$^{1}$\thanks{E-mail: gluscevi@usc.edu}
\\
$^{1}$Department of Physics and Astronomy, University of Southern California, Los Angeles, CA, 90089, USA
}

\date{Accepted XXX. Received YYY; in original form ZZZ}

\pubyear{\the\year{}}

\begin{document}
\label{firstpage}
\pagerange{\pageref{firstpage}--\pageref{lastpage}}
\maketitle

\begin{abstract}
We explore the potential of upcoming 21-cm interferometric observations to probe interacting dark matter (IDM). We focus on scenarios where the dark matter-baryon scattering cross-section scales as $\sigma(v) =\sigma_{0} v^n$, with $\sigma_{0}$ being the normalization constant, $v$ the relative velocity between dark matter and baryons, and $n$ characterizing the velocity dependence. Specifically, we emphasize two cases: Coulomb-like interaction ($n = -4$) and velocity-independent interaction ($n = 0$). Using detailed simulations of the 21-cm power spectrum and the Fisher matrix formalism, we forecast the sensitivity of the Hydrogen Epoch of Reionization Array (HERA), which targets the frequency range 50–225 MHz, to both IDM and astrophysical parameters. We marginalize over key astrophysical uncertainties, including star formation efficiency, ionizing photon escape fraction, and X-ray luminosity. Our results demonstrate that 21-cm power spectrum measurements can significantly improve sensitivity to IDM cross-section, with at least a factor of five improvement over global signal forecasts for the $n=0$ case, and more than an order of magnitude enhancement for the $n=-4$ scenario. These forecasts also significantly improve upon the existing bounds from cosmic microwave background and Milky Way satellite abundance observations. Our analysis also shows that the IDM cross-section exhibits no correlation with the parameters associated with star formation efficiency and ionizing photon escape fraction of Population-II stars. However, we find that the Coulomb-like cross-section is positively correlated with X-ray luminosity. Our results highlight the critical role of accounting for astrophysical uncertainties in obtaining robust inferences of dark matter-baryon interactions from future 21-cm power spectrum observations.
\end{abstract}

\begin{keywords}
cosmology: dark ages, reionization, first stars - cosmology: dark matter - cosmology: theory
\end{keywords}

\section{Introduction}\label{sec:intro}
The existence of dark matter (DM) is well established through diverse cosmological and astrophysical observations, including galaxy rotation curves~\citep{rubin1970rotation, sofue2001rotation}, gravitational lensing~\citep{refregier2003weak, massey2010dark}, galaxy cluster dynamics~\citep{zwicky2009republication, clowe2006direct}, large-scale structure surveys~\citep{tegmark2004cosmological, eisenstein2005detection}, and cosmic microwave background (CMB) anisotropies~\citep{smoot1992structure, bennett2003microwave, aghanim2020planck, hinshaw2013nine}. Despite these observations, the fundamental nature of DM remains elusive, posing one of the most pressing challenges in contemporary cosmology and particle physics. The standard cosmological paradigm describes DM as cold, non-relativistic, and collisionless, commonly referred to as cold dark matter (CDM). However, persistent inconsistencies between CDM predictions and observations on a variety of scales have prompted a critical reassessment of its completeness and motivated the development of alternative DM models (e.g.~\citep{tulin2018dark, abdalla2022cosmology, perivolaropoulos2022challenges, zentner2022critical, schoneberg2022h0, he2023s8}). 

An intriguing extension to the standard CDM framework is the interacting dark matter (IDM) model, which incorporates non-gravitational interactions of DM with baryons. IDM scenarios have been proposed theoretically as extensions to the standard particle physics picture of DM~\citep{sigurdson2004dark, boehm2005constraints, mcdermott2011turning, berlin2018severely, escudero2018fresh}, and have been extensively explored phenomenologically through cosmological observations~\citep{xu2021constraints, becker2021cosmological, boddy2022investigation, li2023atacama}, astrophysical probes such as 21-cm cosmology~\citep{chen2002cosmic, fialkov2018constraining, barkana2018possible, barkana2018signs, munoz2018insights, munoz201821, liu2018implications, liu2019reviving, barkana2023anticipating}, and large-scale structure studies~\citep{short2022dark}. In IDM models, the interaction cross-section depends on the relative velocity between DM and baryons ($v$) and is often modelled as a power-law function: $\sigma(v) = \sigma_0 v^n$, where $\sigma_0$ denotes the amplitude of the cross-section. In this study, we focus on two representative IDM scenarios: Coulomb-like interactions with $n=-4$~\citep{slatyer2018early, boddy2018critical, lin2023dark} and velocity-independent interactions with $n=0$~\citep{boddy2018first, gluscevic2018constraints, nadler2019constraints, nadler2021constraints, maamari2021bounds, rogers2022limits}. IDM can leave observable imprints on a range of cosmological observables across different scales~\citep{gluscevic2019cosmological} and has been extensively studied using the observations of the CMB~\citep{ali2015constraints, nguyen2021observational, gluscevic2018constraints, xu2018probing, slatyer2009cmb}, Lyman $\alpha$ forest~\citep{dvorkin2014constraining, ooba2019cosmological, xu2018probing}, and the abundance of Milky Way satellites~\citep{maamari2021bounds, nadler2021constraints, nguyen2021observational}. In addition to these probes, the cosmological 21-cm signal has emerged as a promising avenue for detecting potential imprints of DM-baryon interactions~\citep{edges_nature, singh2022detection, burns2019dark, adams2023improved, munoz2015heating, boddy2018critical, kovetz2018tighter, mosbech2023probing, sun2025inhomogeneous, flitter202321cmfirstclass_I, flitter202321cmfirstclass_II}. In a recent work, we investigated the sensitivity of current and upcoming global 21-cm signal experiments to IDM models~\citep{global_21}. In this study, we extend that analysis to include spatial fluctuations in the 21-cm signal, focusing on 21-cm power spectrum measurements as a complementary probe of new physics. Several upcoming interferometric experiments, such as HERA~\citep{deboer2017hydrogen, garsden202121}, the Square-Kilometer Array (SKA)~\citep{barry2022ska}, the Murchison Wide- field Array (MWA)~\citep{li2018comparing}, Low Frequency ARray (LoFAR)~\citep{van2013lofar}, Owens Valley Radio Observatory Long Wavelength Array (OVRO-LWA)~\citep{eastwood201921}, and Giant Metrewave Radio Telescope (GMRT)~\citep{chowdhury2021giant}, are specifically designed to measure this observable.

In this work, we forecast the sensitivity of forthcoming 21-cm power spectrum measurements with HERA to DM-baryon interactions and relevant astrophysical parameters. We utilize the Fisher matrix formalism and quantify the precision with which these measurements can forecast the sensitivity of HERA to the interaction cross-section, while marginalizing over degeneracies with various astrophysical processes. Our analysis shows that the 21-cm power spectrum measurements by HERA can significantly improve existing constraints derived from the abundance of the Milky Way satellites~\citep{maamari2021bounds}, the CMB~\citep{nguyen2021observational}, and previous forecasts based on the global 21-cm signal observations~\citep{global_21}. These results highlight the potential of 21-cm power spectrum observations as a powerful and complementary probe of IDM models.

This paper is organized as follows. In Section~\ref{sec:21cmPS}, we briefly review the theoretical framework of the 21-cm power spectrum and outline the modifications introduced by the IDM scenarios. Section~\ref{sec:methods} details our methodology, beginning with the simulation pipeline used to generate the 21-cm power spectrum, followed by the Fisher forecasting formalism, and concluding with a description of the noise models and experimental configurations considered. In Section~\ref{sec:results}, we present our results for the projected constraints on the IDM cross-section and compare them with the existing bounds and forecasts. Finally, in Section~\ref{sec:conclusions}, we summarize the main findings of this study and discuss future prospects.

Throughout this analysis, we adopt the standard cosmological parameters from the Planck 2018 results~\citep{aghanim2020planck}: $h = 0.6736$, $\Omega_\mathrm{m} = 0.3153$, $\Omega_\mathrm{\Lambda} = 0.6847$, $\Omega_\mathrm{b} = 0.04930$, $T_\mathrm{CMB} = 2.72548$, $n_\mathrm{s} = 0.9649$, $N_\mathrm{eff} = 3.046$, and $\sigma_\mathrm{8} = 0.8111$.

\section{21-cm Power Spectrum with IDM}\label{sec:21cmPS}
The redshifted 21-cm signal from neutral hydrogen is commonly characterized by its differential brightness temperature relative to the CMB~\citep{furlanetto2006global, madau1997reionization}
\begin{eqnarray} \label{eq:exact21}
    T_\mathrm{21}(\boldsymbol{x}, z) = \frac{T_S(\boldsymbol{x}, z) - T_\mathrm{\gamma}(z)}{1 + z} \left(1 - e^{-\tau_\mathrm{21}(\boldsymbol{x}, z)}\right)\,,
\end{eqnarray}
where $T_S$, $T_\gamma$, and $\tau_{21}$ denote the spin temperature of neutral hydrogen, CMB temperature at redshift $z$, and the 21-cm optical depth, respectively. The optical depth depends on the local neutral hydrogen density, spin temperature, and the line-of-sight velocity gradient~\citep{field1959, madau1997reionization, furlanetto2006physics}.

The spin temperature, $T_S$, determines the relative population of hydrogen atoms in the hyperfine triplet and singlet states, and is governed by the competition between three processes: coupling to the CMB, collisional coupling with the gas, and resonant scattering of Ly $\alpha$ photons (Wouthuysen–Field effect)~\citep{field1959, Zygelman2005, furlanetto2006physics}. The spin temperature is defined as:
\begin{align} \label{eq:spinT}
    T_\mathrm{S}^{-1} = \frac{T_\mathrm{\gamma}^{-1} + x_\mathrm{c} T_\mathrm{K}^{-1} + x_\mathrm{\alpha} T_\mathrm{c}^{-1}}{1 + x_\mathrm{c} + x_\mathrm{\alpha}},
\end{align}
where $T_\mathrm{K}$ is the kinetic temperature of the gas, $T_\mathrm{c}$ is the color temperature of the Ly $\alpha$ photons, and $x_\mathrm{c}$ and $x_\mathrm{\alpha}$ are the collisional and Ly $\alpha$ coupling coefficients, respectively. The evolution of spin temperature directly impacts the amplitude and sign of the 21-cm signal and is sensitive to astrophysical processes, such as star formation and radiation backgrounds, and potential modifications to the thermal history introduced by DM-baryon interactions.

Interferometric experiments such as HERA are sensitive to spatial fluctuations in the brightness temperature field, $T_{\mathrm{21}}(\boldsymbol{x}, z)$. These fluctuations arise from variations in the baryon density, ionization fraction, spin temperature, and peculiar velocity gradients~\citep{barkana2005probes, furlanetto2006physics}. The statistical characterization of the 21-cm signal at the perturbative level has long been established as a key tool for probing the high-redshift Universe~\citep{madau1997reionization, furlanetto2006physics, pritchard2012constraining, barkana2005probes}. Fluctuations in the differential 21-cm brightness temperature, defined as $\delta T_{21}(\boldsymbol{x}, z) \equiv T_{21}(\boldsymbol{x}, z) - \bar{T}_\mathrm{21}(z)$, capture spatial variations in the 21-cm signal arising from inhomogeneities in the gas density, kinetic temperature, spin temperature, and radiation backgrounds, where $\bar{T}_\mathrm{21}(z)$ denotes the global (spatially averaged) 21-cm brightness temperature at redshift $z$. These perturbations are sensitive to astrophysical processes and potential modifications from DM-baryon interactions. A widely used statistical measure of these fluctuations is the 21-cm power spectrum, defined as
\begin{align} \label{eq:p21_def}
    \langle \tilde{\delta T_\mathrm{21}}(\boldsymbol{k}) \tilde{\delta T_\mathrm{21}}^*(\boldsymbol{k'}) \rangle = (2\pi)^3 \delta_\mathrm{D}(\boldsymbol{k} - \boldsymbol{k'}) P_\mathrm{21}(k)\,,
\end{align}
where $\tilde{\delta T_\mathrm{21}}(\boldsymbol{k})$ denotes the Fourier transform of the spatial fluctuations $\delta T_\mathrm{21}(\boldsymbol{x})$ and $P_\mathrm{21}(k)$ represents the three-dimensional 21-cm power spectrum. This quantity captures the spatial correlations in the signal and provides insights into underlying density fluctuations, astrophysical heating, and reionization processes. The spherically averaged 21-cm power spectrum quantifies the variance of these fluctuations and is given by
\begin{equation}
\Delta_\mathrm{21}^2(k, z) = \frac{k^3}{2\pi^2} P_\mathrm{21}(k, z).
\end{equation}
This observable is the primary target of interferometric experiments such as HERA.

DM-baryon elastic scattering can significantly impact the evolution of the 21-cm power spectrum compared to the standard $\Lambda$CDM cosmology. The primary effects of IDM on the 21-cm power spectrum stem from its influence on spin temperature and the growth of the structure. To understand the IDM cross-section effects on the 21-cm power spectrum, one can model the Universe as discretized into Mpc-scale patches, each with a coherent relative bulk velocity. In the Coulomb-like scenario, the cross-section increases at low relative velocities. DM-baryon scattering can remove heat from the baryons, lowering the gas kinetic temperature, which drives down the spin temperature via collisional and Ly $\alpha$ coupling. Therefore, patches with lower relative velocities cool more efficiently than those with higher relative velocities. This enhanced cooling deepens the absorption feature in the global 21-cm signal and amplifies fluctuations in the power spectrum at frequencies corresponding to the Cosmic Dawn~\citep{fialkov2018constraining}. 

Unlike Coulomb-like scattering, velocity-independent interactions maintain relatively constant efficiency over a broad range of redshifts, cumulatively delaying the thermal and structural evolution of the gas. In this model, scattering between DM and baryons can suppress the small-scale structure formation, similar to warm DM damping, which delays the formation of the first minihaloes. As a result, the formation of galaxies that produce Ly $\alpha$ radiation and X-ray heating is delayed, leading to the postponement of the onset of the 21-cm absorption. Therefore, the IDM model with velocity-independent cross-section causes the peak position of the 21-cm power spectrum to shift towards higher frequencies, with more pronounced shifts occurring for larger interaction cross-sections. 

It is important to note that both low-velocity and very high-velocity regions are rare in the Universe due to their Maxwell-Boltzmann distribution~\citep{2012MNRAS.424.1335F}. Therefore, in Coulomb-like interactions, where IDM effectively targets low relative velocity regions, the cooling effect dominates. On the other hand, in the velocity-independent scenario, the main impact on the 21-cm power spectrum arises from suppressed structure formation, since the IDM in this model affects all velocity regions equally. These distinct signatures on the 21-cm power spectrum motivate us to explore these models further, particularly in the context of upcoming interferometric experiments like HERA. To compute the 21-cm power spectrum within the IDM cosmology, we use the publicly available code \textsc{21cmFirstClass}\footnote{\url{https://github.com/jordanflitter/21cmFirstCLASS}} ~\citep{flitter202321cmfirstclass_I, flitter202321cmfirstclass_II}, which incorporates the effects of IDM by modifying the baryonic temperature and relative bulk velocity evolution in the initial conditions supplied to \textsc{21cmFAST}~\citep{mesinger201121cmfast, munoz2022impact}. 

The scale and timing of the deviations induced by DM-baryon interactions in the 21-cm power spectrum depend sensitively on interaction strength ($\sigma_0$), the DM mass ($m_\mathrm{\chi}$), and key astrophysical parameters that govern the thermal and ionization history of the IGM. In our analysis, we consider a set of astrophysical parameters, including the star formation efficiency ($f_\mathrm{\ast}$), the slope of the stellar-to-halo mass relation ($\alpha_\mathrm{\ast}$), the escape fraction of ionizing photons ($f_{\mathrm{esc}}$), the slope of the X-ray spectral energy distribution ($\alpha_{\mathrm{esc}}$), and the X-ray luminosity per unit star formation rate ($L_\mathrm{X}$). These parameters collectively influence the timing and efficiency of Ly $\alpha$ coupling, X-ray heating, and reionization, thereby shaping the 21-cm power spectrum.

\begin{figure}
	\includegraphics[scale=0.4]{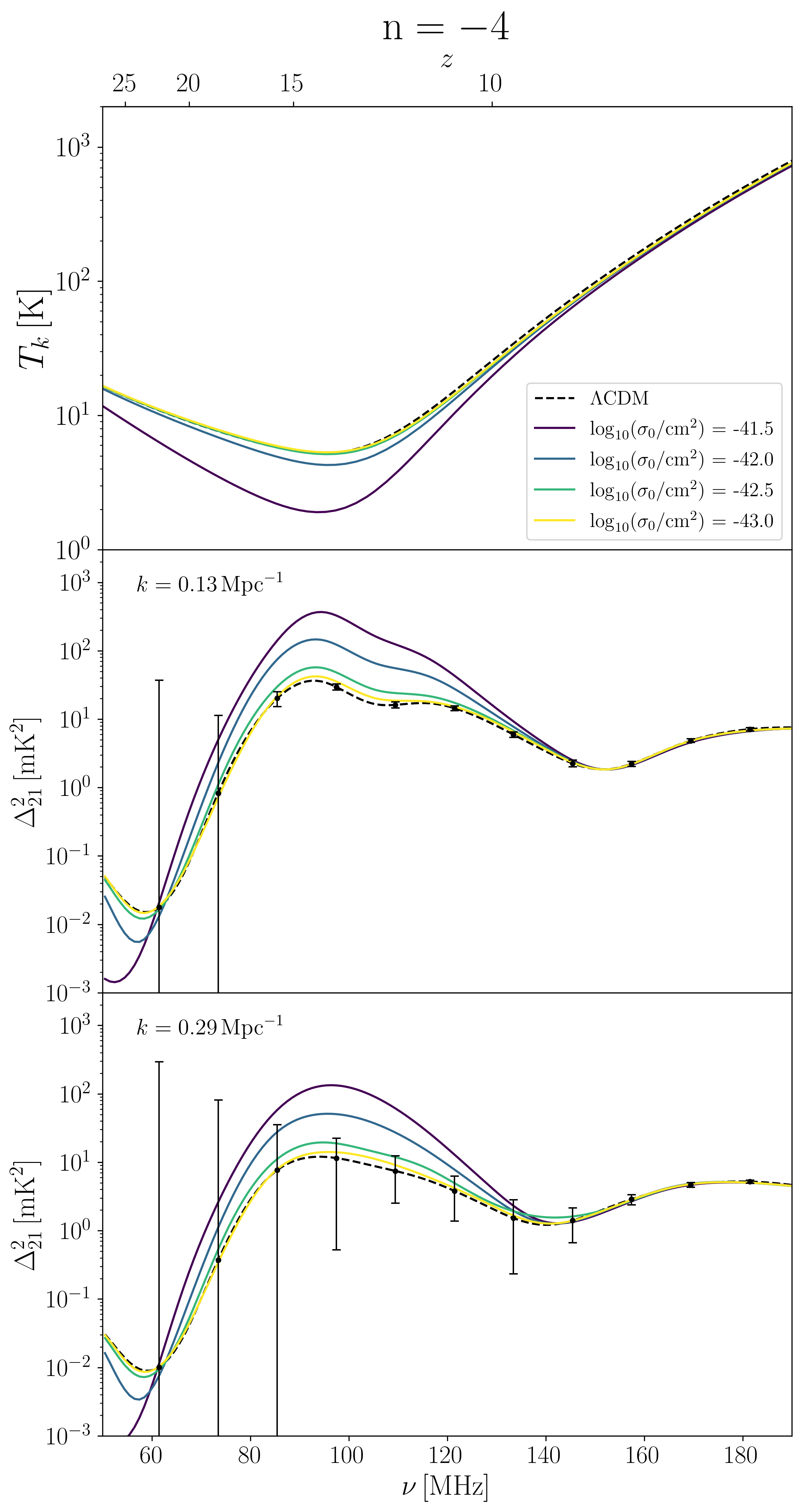}
	\centering
	\caption{
		The top panel illustrates the evolution of the gas kinetic temperature as a function of observed frequency $\nu$ for an IDM model with Coulomb-like interaction at fixed DM mass \(m_{\chi} = 1~\mathrm{GeV}\). Solid curves correspond to different cross-section amplitudes \(\sigma_0\) for IDM, while the dashed curve represents the standard \(\Lambda\)CDM model. The middle and bottom panels show the 21-cm power spectrum $\Delta^{2}_{21}(k)$ as a function of frequency at \(k = 0.13,\,0.29~\mathrm{Mpc}^{-1}\). The error bars indicate the uncertainties included in the forecast for HERA, under the "moderate noise" scenario, described in the text.}
	\label{plt:ps_n-4_2k}
\end{figure}

\begin{figure}
	\includegraphics[scale=0.4]{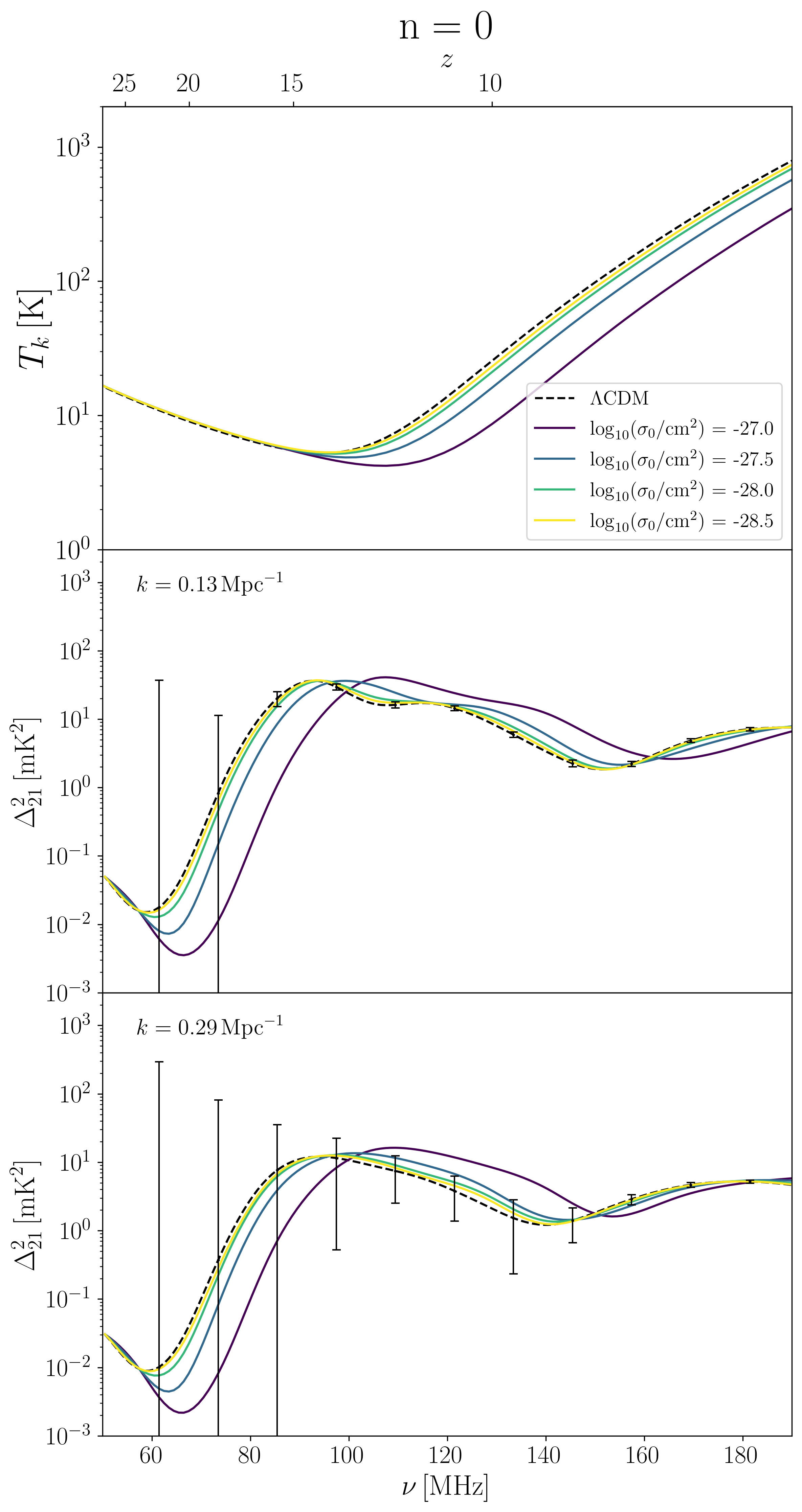}%
	\centering
	\caption{ 
		The top panel illustrates the evolution of the gas kinetic temperature as a function of observed frequency $\nu$ for an IDM model with velocity-independent interaction at fixed DM mass \(m_{\chi} = 1~\mathrm{GeV}\). Solid curves correspond to varying cross-section amplitudes \(\sigma_0\) for IDM, while the dashed curve represents the standard \(\Lambda\)CDM model. The middle and bottom panels show the 21-cm power spectrum $\Delta^{2}_{21}(k)$ as a function of frequency at \(k = 0.13,\,0.29~\mathrm{Mpc}^{-1}\). The error bars indicate the uncertainties included in the forecast for HERA, under the "moderate noise" scenario, described in the text.}
	\label{plt:ps_n0_2k}
\end{figure}

In the standard $\Lambda$CDM scenario, the 21-cm power spectrum evolves with redshift, as different physical processes become relevant at different times; see Figs.~\ref{plt:ps_n-4_2k} and~\ref{plt:ps_n0_2k} for illustration. At early times ($z \gtrsim 20$, $\nu \lesssim 75~\mathrm{MHz}$), the rise in the power is driven by the increase in spatial fluctuations of the Ly$\alpha$ coupling $x_\alpha$, as the first stars begin to form. As X-ray sources begin heating the IGM, patchy temperature fluctuations introduce another distinct peak in the power spectrum, around $120~\mathrm{MHz}$, with the exact amplitude and position depending on astrophysical parameters such as the X-ray efficiency. At later times, as reionization progresses ($z \lesssim 10$, $\nu \gtrsim 130~\mathrm{MHz}$), the contrast between ionized and neutral regions dominates the signal; the power gradually diminishes as the neutral hydrogen fraction approaches zero. Across these epochs, the power spectrum also exhibits scale dependence: larger scales ($k \sim 0.1~\mathrm{Mpc}^{-1}$) display stronger fluctuations due to their sensitivity to large-scale structure and heating inhomogeneities, while smaller scales are more rapidly smoothed out as the IGM becomes ionized and heated on those scales.

Figs.~\ref{plt:ps_n-4_2k} and~\ref{plt:ps_n0_2k} illustrate the impact of DM-baryon interactions on the gas temperature (top panels) and the 21-cm power spectrum (lower two panels) for the Coulomb-like and velocity-independent IDM scenarios, respectively. These results correspond to a fixed DM mass of $m_\mathrm{\chi} = 1~\mathrm{GeV}$ and different wavenumbers, e.g., $k = 0.13$ and $0.29~\mathrm{Mpc}^{-1}$. The sensitivity of the power spectrum to the interaction strength arises from its dependence on the gas temperature and the IDM's impact on the growth of the structure. As explained earlier in this section, for a Coulomb-like scattering scenario, a higher interaction strength results in a greater contrast between gas temperatures in hot and cold patches, thus amplifying the power, particularly within the frequency range corresponding to the Cosmic Dawn ($\sim80$-$140~\mathrm{MHz}$), which is evident in Fig.~\ref{plt:ps_n-4_2k}. However, despite the strong impact of Coulomb-like interactions on gas temperature and small-scale fluctuation amplitude, this scenario does not result in a significant delay of astrophysical milestones, such as the formation of the first stars and the onset of X-ray heating, as scattering is the most effective at Cosmic Dawn. Hence, we observe only an amplitude enhancement in the 21-cm power spectrum without any significant shift in peak position as clearly evident from Fig.~\ref{plt:ps_n-4_2k}. This result is consistent with the findings reported in the original 21cmFirstCLASS calculations for $n=-4$~\citep{flitter202321cmfirstclass_I}. 
Additionally, we can also see in Fig.~\ref{plt:ps_n-4_2k} that the power spectrum is suppressed compared to the standard cosmology for $\nu\lesssim 60$MHz. On the other hand, Fig.~\ref{plt:ps_n0_2k} clearly shows that the velocity-independent IDM mainly shifts the power spectrum peaks toward higher frequencies, with larger cross-sections producing more pronounced shifts. Such shifts in peak positions lead to a cross-over of curves as visible in the figure, making it distinguishable from the velocity-dependent case. As explained earlier, this shift arises mainly from the suppression of structure formation, which is the dominant effect of velocity-independent interactions.

\section{Methods}\label{sec:methods}
In this work, we study the impact of IDM on the 21-cm power spectrum and assess the capability of HERA to distinguish between the 21-cm power spectra predicted by IDM and standard $\Lambda$CDM models. Using the Fisher matrix formalism with a $\Lambda$CDM fiducial cosmology, we forecast the expected sensitivity on IDM cross-section and relevant astrophysical parameters. In the following sections, we briefly review the theoretical framework of the 21-cm power spectrum, present the Fisher matrix formalism tailored for 21-cm signal forecasting, and summarize the noise models and experimental configurations used in our analysis.

\subsection{21-cm simulations}\label{subsec:21-cm sims}
We use the publicly available semi-numerical code, \textsc{21cmFirstCLASS}~\citep{flitter202321cmfirstclass_I, flitter202321cmfirstclass_II}, which utilizes initial conditions from a modified version of \textsc{CLASS}\footnote{\url{https://github.com/kboddy/class_public/tree/dmeff}}~\citep{gluscevic2018constraints, boddy2018first, nguyen2021observational} as input for \textsc{21cmFAST}~\citep{mesinger201121cmfast, munoz2022impact} to generate the 21-cm power spectrum in the IDM regime. This framework enables a consistent treatment of the 21-cm signal in cosmologies beyond $\Lambda$CDM, including models with IDM. It simulates the spatial fluctuations in the 21-cm brightness temperature while accounting for IDM-induced modifications to structure formation and thermal history. The code first computes the relevant initial conditions, such as the matter density fluctuations, relative velocity transfer functions, growth rates, and background properties like the kinetic temperature and ionization fraction, and subsequently evolves these quantities from recombination through Cosmic Dawn. This approach allows for a physically consistent simulation of the 21-cm power spectrum in the presence of IDM.

In this study, we focus on scenarios in which DM interacts with baryons~\citep{dvorkin2014constraining, munoz2015heating, boddy2018critical, fialkov2018constraining, xu2018probing, barkana2018possible, short2022dark, he2023s8, driskell2022structure, fischer2025n}, resulting in momentum and heat exchange. In this model, the total scattering target energy density is taken to be the total baryonic energy density, $\rho_\mathrm{target} = \rho_b$, and the effective target mass is set to the mean baryon mass, $m_t = \bar{m}_b$. The average baryon mass incorporates both neutral hydrogen and helium contributions and is given by
\begin{equation}
    \bar{m}^{-1}_b = \left( (1-Y_{\mathrm{He}}) \ m^{-1}_{\mathrm{H}} + Y_{\mathrm{He}} \ m^{-1}_{\mathrm{He}} \right) \ (1 - x_e),
\end{equation}
where $m_{\mathrm{H}}$ and $m_{\mathrm{He}}$ are the masses of hydrogen and helium atoms respectively, $Y_{\mathrm{He}} \approx 0.245$ is the helium mass fraction, and $x_e$ is the free electron fraction. 

Our simulations are performed in a $1024~\mathrm{Mpc}$ wide computational box, discretized into grid cells with a spatial resolution of $128~\mathrm{Mpc}$. The chosen box size ensures adequate statistical sampling at large scales ($k \gtrsim 0.1~\mathrm{Mpc}^{-1}$), where the 21-cm signal exhibits the most power. In parallel, the grid cell resolution of approximately $1~\mathrm{Mpc}$ provides sufficient accuracy for the power spectrum to converge at smaller scales ($k \approx 0.5~\mathrm{Mpc}^{-1}$), which aligns well with the sensitivity range of HERA. To ensure the robustness of our results, we performed convergence tests and verified that this configuration yields stable power spectra across the relevant $k$-range. We also confirmed that the box size is sufficiently large to avoid the effects of cosmic variance, particularly for IDM scenarios where large-scale modes are more sensitive to the altered thermal and density evolution. In particular, we verified that the level of cosmic variance in our simulations remains well below the projected HERA error bars across the full range of scales probed, ensuring that our forecasts are not limited by sample variance. These findings are consistent with earlier convergence studies using \textsc{21cmFAST} and \textsc{21cmFirstCLASS}~\citep{sun2025inhomogeneous, adi2025early}.

\subsection{Fisher analysis}\label{subsec:fisher}
To quantify the constraining power of current and future 21-cm power spectrum experiments on IDM and astrophysical parameters, we employ the Fisher information matrix formalism. This framework approximates the likelihood surface near a fiducial model as Gaussian and provides a computationally efficient method to estimate parameter uncertainties. Specifically, we adopt a formulation of the Fisher formalism tailored to the 21-cm power spectrum, where the observable is the spherically averaged dimensionless power spectrum, $\Delta_\mathrm{21}^2(\mathrm{k,z})$, evaluated as a function of wavenumber $k$ and redshift $z$~\citep{mason202321cmfish}:
\begin{eqnarray} \label{eq:fisher}
    F_{\mathrm{ij}} = \sum_{\mathrm{k,z}} \frac{1}{[\delta \Delta_{21}^2(\mathrm{k,z})]^{2}} \frac{\partial \Delta_\mathrm{21}^{2}(\mathrm{k,z})}{\partial \theta_\mathrm{i}} \frac{\partial \Delta_\mathrm{21}^{2}(\mathrm{k,z})}{\partial \theta_\mathrm{j}}\,,
\end{eqnarray}
where $\delta \Delta_\mathrm{21}^2(k,z)$ represents the uncertainty in the measured 21-cm power spectrum at a given wavenumber $k$ and redshift $z$, and $\theta_\mathrm{i}$ represents the set of astrophysical and IDM parameters used in this analysis 
\begin{align}
     \theta_\mathrm{i} \in \ \{ & f_\mathrm{\ast}^{(\text{II})}, \ \alpha_\mathrm{\ast}^{(\text{II})}, \ f_\mathrm{esc}^{(\text{II})}, \ \alpha_\mathrm{esc}^{(\text{II})}, \ L_\mathrm{X}^{(\text{II})}, \nonumber \\
     & m_\mathrm{\chi}, \ \sigma_\mathrm{0} \}\,.
\end{align}
The set of astrophysical parameters includes the star formation efficiency \( f_\mathrm{\ast}^{(\text{II})} \) for Population-II stars, normalized by the star formation rate (SFR), and the power law index \( \alpha_\mathrm{\ast}^{(\text{II})} \), which describes the dependence of star formation on halo mass. We also consider the escape fraction \( f_{\text{esc}}^{(\text{II})} \), representing the fraction of photons produced by Population-II stars that escape their host galaxies and ionize the IGM, as well as the power law index \( \alpha_\mathrm{\text{esc}}^{(\text{II})} \) that governs this escape fraction. The X-ray luminosity \( L_\mathrm{X}^{(\text{II})} \) of Population-II stars, which influences the ionization process in the early Universe, is another key parameter. In addition to these astrophysical parameters, we include the IDM parameters \( m_\mathrm{\chi} \) and \( \sigma_\mathrm{\text{0}} \), which correspond to the IDM mass and interaction cross-section, respectively. These parameters play a crucial role in modulating the effects of DM-baryon interactions on the 21-cm signal. For further details and physical motivations behind the astrophysical parameters, we refer the reader to \citet{mesinger201121cmfast}. The fiducial values of both astrophysical and IDM parameters used in this study are listed in Table~\ref{tab:fid_vals}.
While our analysis explicitly marginalizes over only the astrophysical parameters associated with Population-II star formation, we have fixed the parameters governing Population-III stars at their default fiducial values. Given the sensitivity of Population-III stars to DM-baryon relative velocities~\citep{hirano2018baryon}, explicitly marginalizing over Population-III parameters (such as $L_{X}^{\mathrm{III}}$, $\alpha_{\mathrm{esc}}^{\mathrm{III}}$, $\alpha_{\star}^{\mathrm{III}}$, $f_{\mathrm{esc}}^{\mathrm{III}}$, and $f_{\star}^{\mathrm{III}}$) could impact our results. Recent forecasts by \citet{adi2025early}, though focused on Early Dark Energy, demonstrate non-negligible but not strong degeneracies between Population-II and Population-III parameters. It is possible that marginalizing the Population-III parameters could potentially broaden the uncertainties and relax our constraints. However, a complete exploration of these effects is beyond the scope of this work and is deferred to a future study.

\begin{table}
		\centering
	\resizebox{0.6\linewidth}{!}{%
	\begin{tabular}{|c|c|}
		\hline
		Parameter & Fiducial value\\
		\hline
		$\log_{10} f_\mathrm{\ast}^{(\text{II})}$ & -1.25 \\
		$\alpha_\mathrm{\ast}^{(\text{II})}$ & 0.5 \\
		$\log_{10} f_\mathrm{esc}^{(\text{II})}$ & -1.35\\
		$\alpha_\mathrm{esc}^{(\text{II})}$ & -0.3\\
		$L_\mathrm{X}^{(\text{II})}$ & 40.5\\
		$m_\mathrm{\chi}$ & 100 kev-1 Tev\\
		$\sigma_\mathrm{0}$ & 0.0 (CDM)\\
		\hline
	\end{tabular}}
	\caption{Fiducial values of the IDM and astrophysical parameters used in the Fisher matrix analysis. These values are adopted from~\citet{munoz2022impact} and serve as a representative baseline for forecasting the sensitivity of upcoming 21-cm power spectrum measurements.}
	\label{tab:fid_vals}
\end{table}

\begin{figure*}
	\includegraphics[scale=0.6]{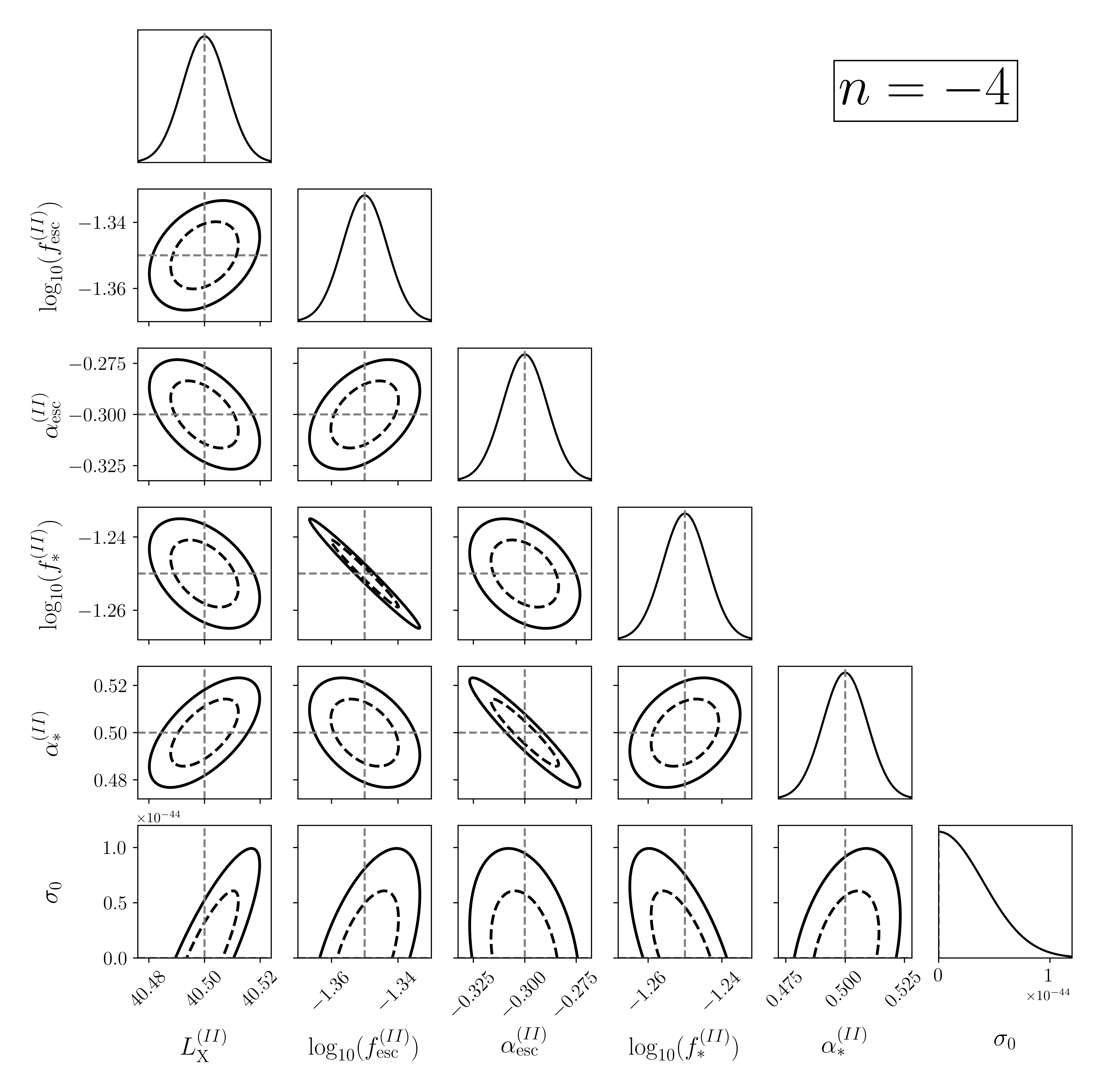}%
	\caption{Forecasted constraints on astrophysical parameters and IDM cross-section for the $n=-4$ model with a fixed DM mass of $10$ MeV, obtained under an optimistic observational configuration, are presented. The $2$D contours represent the forecasted $68\%$ (dotted black) and $95\%$ (solid black) uncertainty regions, while the top panel of each column represents the marginalized $1$D posterior probability distributions. The grey lines indicate the fiducial values of each parameter.}
	\label{plt:n-4_tri}
\end{figure*}

\begin{figure*}
	\includegraphics[scale=0.6]{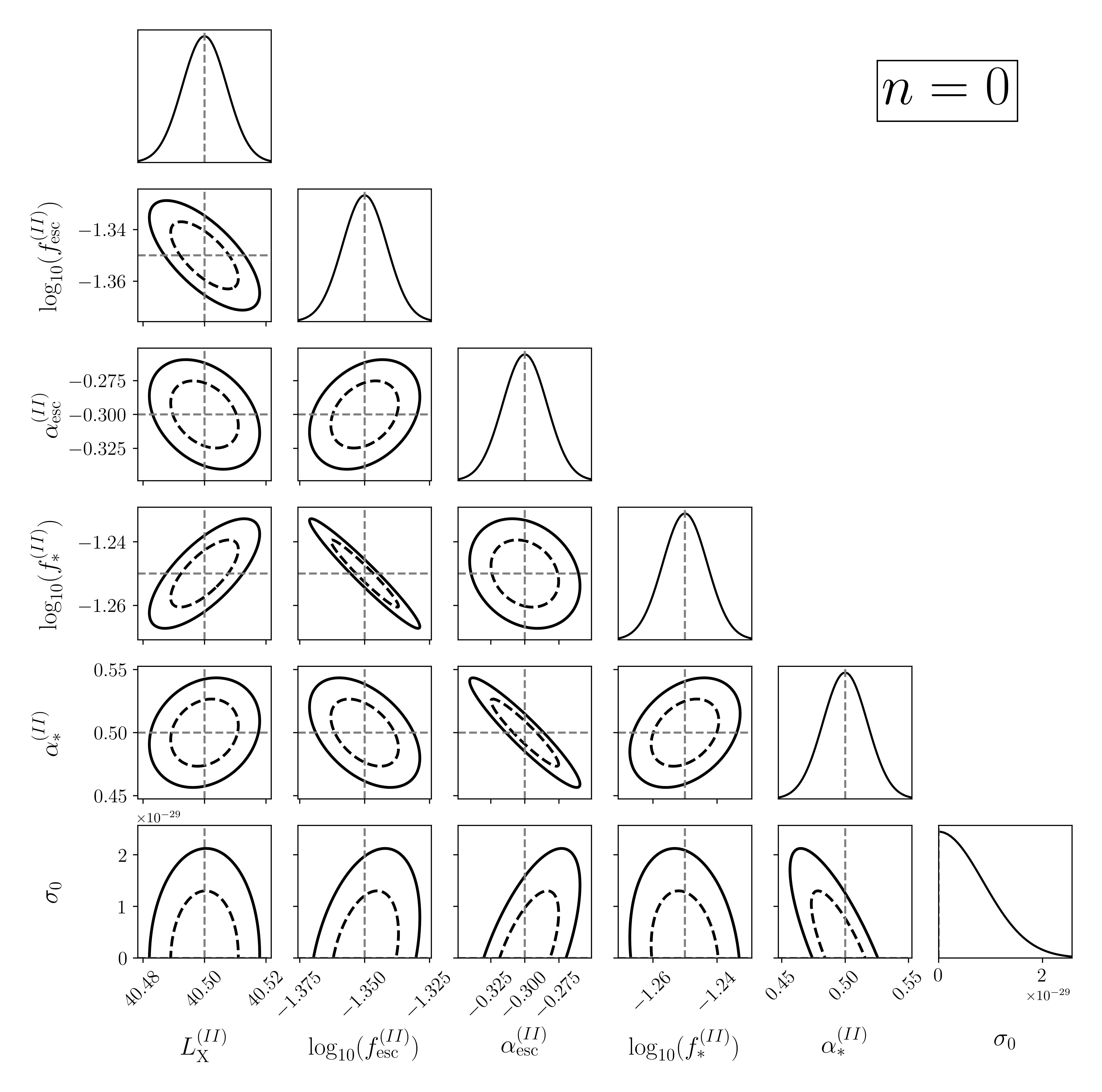}%
	\caption{Forecasts on astrophysical parameters and the cross-section of interactions for the $n=0$ model with a fixed DM mass of $1$ GeV are presented. These forecasts are obtained under an optimistic observational configuration. The $2$D contours represent the $68\%$ and $95\%$ confidence regions, shown by the dotted and solid black curves, respectively, while the top panel of each column represents the marginalized posterior distributions for the corresponding parameters. The grey lines correspond to the fiducial values of each parameter.}
	\label{plt:n0_tri}
\end{figure*}

\subsection{Input covariance noise model}\label{subsec:noise}
To determine the uncertainty in the 21-cm power spectrum, denoted by $ \delta \Delta_\mathrm{21}(\mathrm{k,z}) $ in Eq.~\ref{eq:fisher}, we utilize the~\textsc{21cmSense}\footnote{\url{https://github.com/rasg-affiliates/21cmSense}} \citep{pober2013baryon, pober2014next} code to simulate the expected noise for HERA. The final HERA configuration comprises $331$ antennas, each $14$ meters wide and arranged in a hexagonal pattern, designed to observe in a frequency range of $50-225$ MHz, corresponding to $ z \approx 5-27 $, with an $8$ MHz bandwidth. The analysis is performed across $12$ redshift bins and $5$ wavenumbers. Each frequency band consists of $82$ channels, with a total of $1024$ channels spanning $100$ MHz.

We assume that HERA operates for $540$ days, with $6$ hours of integration time per night. The receiver temperature is set to $100 \ \mathrm{K}$, and the sky temperature is modelled as a frequency-dependent power law
\begin{eqnarray}
    T_\mathrm{\text{sky}}(\nu) = 60 \, \text{K} \, \left( \frac{\nu}{300 \, \text{MHz}} \right)^{-2.55}\,. 
\end{eqnarray}
The total system temperature employed in estimating the thermal noise is the sum of the receiver and sky temperatures. These values are input into the \textsc{21cmSense} pipeline to compute the expected uncertainties in the 21-cm power spectrum across $k$ and $z$ bins, incorporating instrumental layout, observing time, and beam geometry.

To account for foreground contamination, we adopt two foreground removal scenarios as defined within the \textsc{21cmSense} framework: the "moderate" and "optimistic" cases. In the moderate scenario, foregrounds are assumed to occupy a "wedge" in Fourier space, extending beyond the horizon by a fixed buffer angle-specifically, $0.1$ radians beyond the geometric horizon limit. This conservative approach masks all contaminated $k$-modes within the wedge. In the optimistic scenario, foreground removal is assumed to be more effective, excluding only those modes strictly within the primary beam field of view, thereby recovering a larger portion of Fourier space for cosmological analysis. These assumptions directly affect the number of usable $k$-modes at each redshift, hence the achievable sensitivity to the 21-cm power spectrum. For a detailed description of the foreground wedge, buffer angles, and their implementation within the \textsc{21cmSense} framework, we refer the reader to~\citet{pober2014next}.

\begin{figure*}
	\includegraphics[scale=0.65]{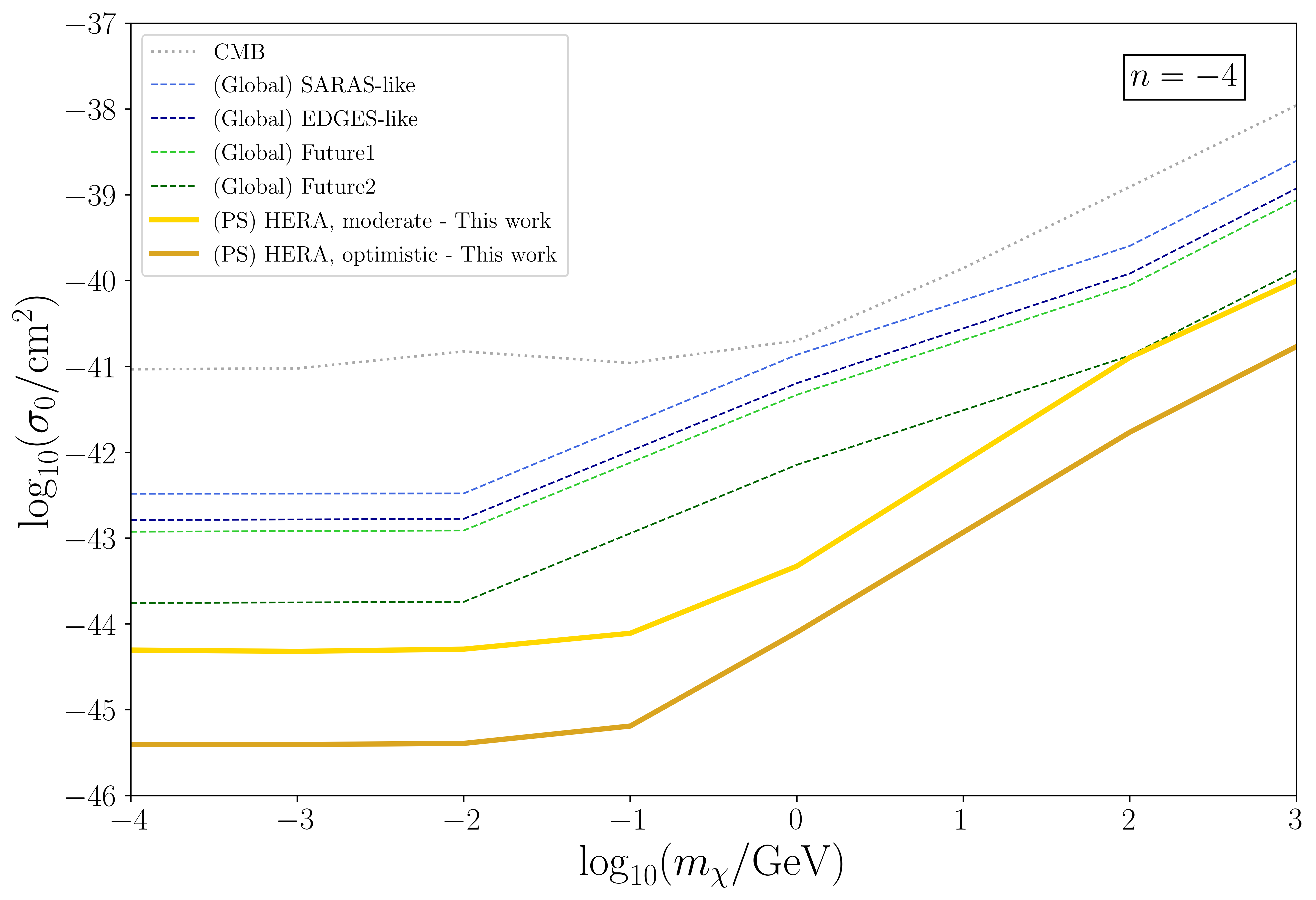}%
	\caption{The 95\% confidence level upper limit forecasts on $\log_{10}(\sigma_{0}/\mathrm{cm}^2)$ as a function of IDM mass are shown for a Coulomb-like model and two different experimental scenarios. The previous bounds from CMB~\citep{nguyen2021observational} and forecasts from the Global 21-cm analysis~\citep{global_21} are also plotted for comparison. Both experimental configurations predict stronger forecasts compared to the current CMB bounds and Global study forecasts.}
	\label{plt:n-4forecasts}
\end{figure*}

\begin{figure*}
	\includegraphics[scale=0.65]{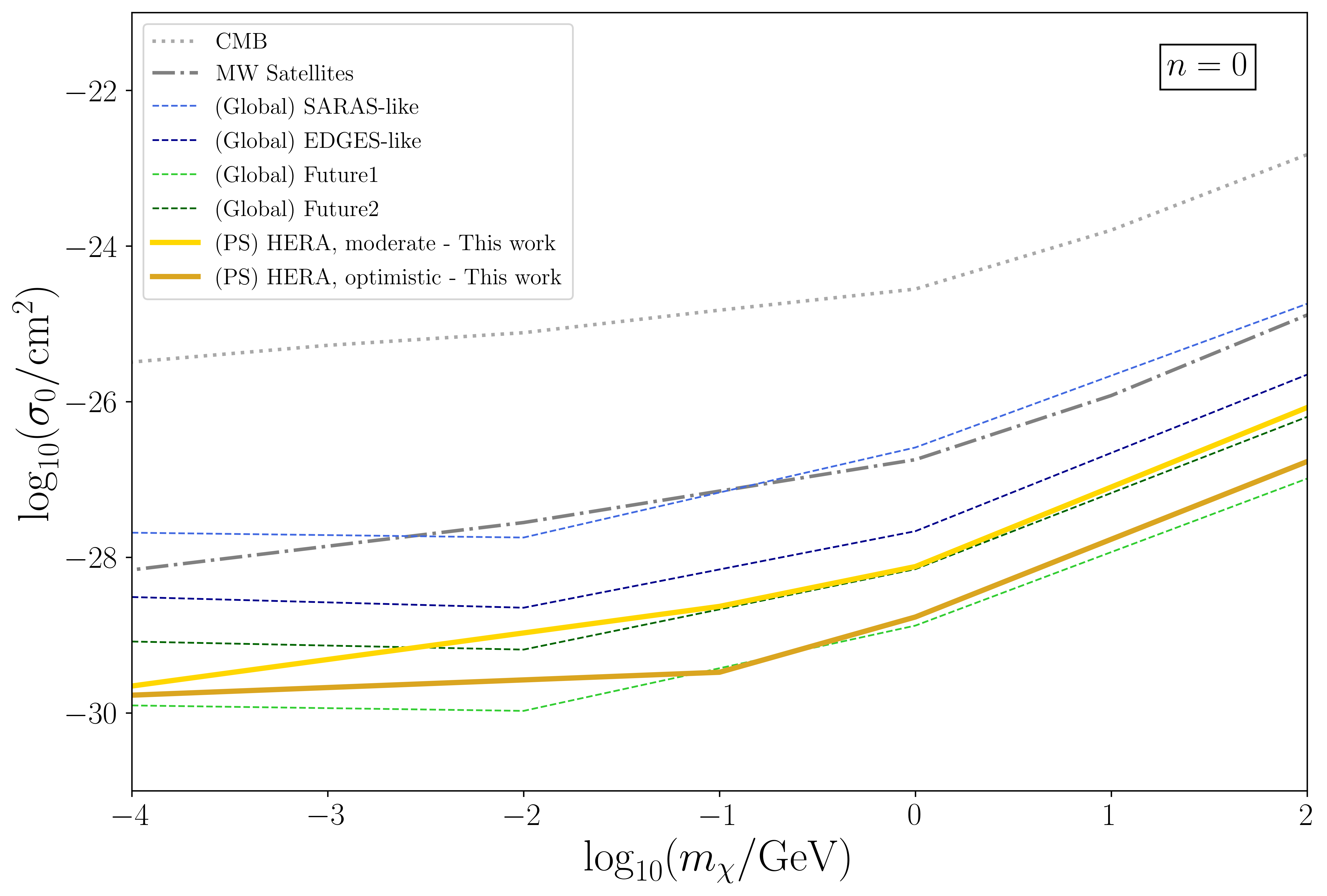} 
	\caption{The 95\% confidence level upper limit forecasts on $\log_{10}(\sigma_{0}/\mathrm{cm}^2)$ as a function of IDM mass are shown for a velocity-independent cross-section IDM model and two different experimental scenarios. The previous bounds from CMB~\citep{nguyen2021observational}, Milky Way satellite abundance~\citep{maamari2021bounds}, and the forecasts from Global 21-cm analysis~\citep{global_21} are also plotted for comparison. Both experimental scenarios predict stronger forecasts compared to the current CMB bounds, Milky Way bounds, and Global study forecasts for EDGES-like and SARAS-like scenarios. However, the forecasts are comparable to the ones from futuristic global signal scenarios.}
	\label{plt:n0forecasts}
\end{figure*}

\section{Results} \label{sec:results}
We generate mock signals with $\Lambda$CDM as our fiducial model, and evaluate the sensitivity of HERA measurements of the 21-cm power spectrum to recovering DM–baryon scattering cross-section, for two interaction models. The fiducial values for each model parameter considered are listed in Table~\ref{tab:fid_vals}. We use \textsc{fishchips} code\footnote{\url{https://github.com/xzackli/fishchips-public}}~\citep{li2018disentangling} to visualize the uncertainties and correlations between parameters as 1-D and 2-D posterior probability distributions shown in Figs.~\ref{plt:n0_tri} and \ref{plt:n-4_tri}. 

Figs.~\ref{plt:n-4_tri} and~\ref{plt:n0_tri} show the projected 68$\%$ and 95$\%$ confidence contours for Coulomb-like and velocity-independent interactions, respectively, assuming an optimistic experimental set-up. In both models, $\sigma_0$ exhibits negligible correlations with $\alpha_{\ast}^{(II)}$ and $\alpha_{\mathrm{esc}}^{(II)}$, due to the fact that variations in the power-law indices governing the halo mass dependence of the escape fraction and star formation efficiency, have a comparatively subtle influence on the 21-cm power spectrum shape in the presence of DM-baryon interactions. The same argument is valid for star formation efficiency and escape fraction coefficients. For the $n=-4$ model, there is a moderate positive correlation between $\sigma_0$ and $\mathrm{L_X}$, which arises because increasing the cross-section enhances baryon cooling and delays IGM heating, thereby requiring higher X-ray luminosities to compensate and maintain the fiducial power spectrum signal. In contrast to the Coulomb-like interaction case, $\sigma_0$ shows no correlation with $\mathrm{L_X}$, indicating their relatively independent roles in shaping the 21-cm power spectrum in case of the $n=0$ model. Additionally, we found the expected strong anti-correlations between $\alpha_{\ast}^{(II)}$ and $\alpha_{\mathrm{esc}}^{(II)}$ as well as between $\mathrm{f}_{\ast}^{(II)}$ and $\mathrm{f}_{\mathrm{esc}}^{(II)}$, which highlights the fact that more massive galaxies tend to have higher star formation efficiencies and lower escape fraction of ionizing photons. We note that the forecasted contours shown in these figures are calculated using the Fisher matrix formalism, which assumes a Gaussian likelihood and thus yields elliptical confidence regions. A complete Markov chain Monte Carlo (MCMC) treatment could reveal non-Gaussian structures or more complex degeneracies that this approach cannot capture.

We repeat our Fisher forecast for a range of DM particle masses and evaluate the $95\%$ confidence-level upper bound on interaction cross-section for each particle mass. The resulting parameter uncertainties are shown in Figs.~\ref{plt:n-4forecasts} and ~\ref{plt:n0forecasts} for Coulomb-like and velocity-independent interaction models, respectively, derived under moderate and optimistic experimental set-ups of HERA. The corresponding forecasts are also presented in Table~\ref{tab:final_n0,-4}. For comparison, we include previous constraints from Planck CMB observations (dotted grey)~\citep{nguyen2021observational} and Milky Way satellite abundance (dot-dashed grey lines)~\citep{maamari2021bounds}, as well as forecasts from global 21-cm signal experiments including Shaped Antenna measurement of the background RAdio Spectrum (SARAS)-like (light blue), Experiment to Detect the Global EoR Signature (EDGES)-like (dark blue), Future1 (light green), and Future2 (dark green)\footnote{For a complete description of these global 21-cm experimental scenarios and their assumptions, see~\citet{global_21}.}. For the $n=-4$ case, we find that HERA can significantly improve upon both existing constraints and also upon the sensitivity of the global signal to the same physics. In particular, the optimistic HERA scenario extends sensitivity by several orders of magnitude compared to global signal experiments. Similarly, for $n=0$ case, HERA forecasts substantially improve upon current observational constraints and forecasts for SARAS-like and EDGES-like experimental configurations. The improvement in sensitivity is especially pronounced in the Coulomb-like interaction model, where the interaction leads to stronger modifications in the thermal evolution and a more significant amplification of the 21-cm power spectrum. The resulting power spectrum exhibits higher contrast features compared to the velocity-independent IDM case, particularly during the heating epoch, making it more detectable. This is partly due to the increased power generated by the Coulomb-like scattering at earlier times, which enhances the sensitivity of interferometric measurements to lower cross-section values. The global‐signal constraints are taken from our companion paper on the forecasts from the global 21-cm signal ~\citep{global_21}, where a fixed foreground model was assumed without marginalizing over foreground parameters. The global 21-cm study assumes extended frequency coverage and large integration times for global signal experiments, which leads to tighter forecasts, compared to constraints from CMB anisotropies and Milky Way satellite abundances. We note that including foreground marginalization as described in~\citet{munoz2021cosmic} may change these forecasts of the global study.

\begin{table}
	\resizebox{0.8\linewidth}{!}{%
	\centering
	\begin{tabular}{|c|c|c|c|}
		\hline
		$n$ & $m_\mathrm{\chi}$ & Moderate & Optimistic\\
		\hline
		& 100 kev & -29.65 & -29.77\\
		& 100 Mev & -28.63 & -29.48\\
		0 & 1 Gev   & -28.11 & -28.77\\
		& 100 Gev & -26.07 & -26.77\\
		\hline
		& 100 kev & -44.30 & -45.41\\
		& 10 Mev  & -44.11 & -45.19\\
		-4 & 1 Gev   & -43.33 & -44.10\\
		& 100 Gev & -40.90 & -41.77\\
		& 1 Tev   & -40.00 & -40.77\\
		\hline
	\end{tabular}}
	\caption{The 95\% confidence level upper limit forecasts on $\log_\mathrm{10}(\sigma_\mathrm{0}/ \mathrm{cm}^{2})$, the coefficient of the momentum-transfer cross-section of DM–baryon scattering, are shown for both the velocity-independent and Coulomb-like models, across a range of IDM masses and two different experimental configurations for HERA.}
	\label{tab:final_n0,-4}
\end{table}

Additionally, we would also like to highlight the effect of the secondary term, which can arise in these models, on our results. One secondary effect worth discussing is the frictional heating introduced by DM-baryon scattering, as described by \citet{munoz2015heating}. This heating arises from the damping of relative bulk velocity fluctuations between DM and baryons, leading to energy dissipation and an additional heating term proportional to $V_{\chi b}^2$. While the implementation in \texttt{21cmFirstCLASS} includes the drag-induced thermal coupling term (proportional to $T_{\chi} - T_b$), it does not account for this frictional heating. This effect is expected to become increasingly important at higher redshifts, for low DM masses ($m_\chi \lesssim \mathrm{few~GeV}$), or large momentum-transfer cross-sections ($\sigma_0 \gtrsim 10^{-41}~\mathrm{cm}^2$), where the relative velocity is more efficiently damped. Including this additional heating could partially counteract the drag-induced cooling in the $n = -4$ case, reducing the amplitude of temperature fluctuations and mildly broadening the contours in the $\sigma_0 - m_\chi$ parameter space. In turn, the power spectrum would exhibit reduced sensitivity to $\sigma_0$, and the projected upper limits may become less stringent in the regions where cooling is currently dominant. However, a full quantitative exploration of this effect is beyond the scope of this work and is left to future work.

\section{Conclusions}\label{sec:conclusions}
Understanding the fundamental nature of DM remains one of the most compelling challenges in modern cosmology. This work explores IDM in light of upcoming 21-cm power spectrum measurements by interferometric experiments like HERA. Utilizing the Fisher forecast formalism, we study the potential of these measurements to forecast the sensitivity of HERA to the DM-baryon interaction cross-section, focusing on two classes of IDM models: Coulomb-like and velocity-independent interactions. These interaction scenarios modify the IGM thermal history in distinct ways, leaving observable imprints on the 21-cm power spectrum.

We performed Fisher matrix forecast analysis tailored to HERA, incorporating degeneracies with key astrophysical parameters, to quantify the projected sensitivity to IDM. Our analysis shows that these measurements can place significantly stronger forecasts on the DM-baryon interaction cross-section, improving upon the existing bounds from CMB and Milky Way satellite abundances. Specifically, for the $n=-4$ case, the forecasts of the upper limits show improvement over current CMB limits~\citep{nguyen2021observational} by at least two orders of magnitude, while in the $n=0$ case, they surpass the current constraints from Milky Way satellite abundances~\citep{maamari2021bounds} by over an order of magnitude. Moreover, our results show an improvement of at least a factor of five over global signal forecasts for EDGES-like experiment~\citep{global_21} for the $n = 0$ case, and more than an order of magnitude improvement for the $n = -4$ scenario, under moderate observational assumptions for HERA. The improvements are further enhanced under an optimistic observation set-up of HERA. These gains in HERA sensitivity stem from the ability of the 21-cm power spectrum to capture the spatial and redshift evolution of the signal, enabling more precise disentanglement between $\Lambda$CDM and IDM. Under the assumption that Population-III parameters are fixed, Population-II stars do not exhibit strong degeneracies with $\sigma_0$, allowing the imprint of DM-baryon interactions on the 21-cm power spectrum to be reconstructed without significant bias from Population-II astrophysics. The only exception to this is the X-ray luminosity, which is negatively correlated with the cross-section of interaction only in the Coulomb-like case. We note, however, that fully marginalizing over Population-III parameters, particularly relevant during Cosmic Dawn, could introduce additional degeneracies and broaden the resulting constraints.

The methodology developed in this work can be extended to incorporate additional sources of uncertainty, such as foreground residuals and instrumental systematics. Future work could also include additional IDM models with different velocity dependence or mass scales, and incorporate real observational data as upcoming 21-cm experiments mature. Our results highlight the power of 21-cm cosmology as a precision probe of beyond the standard CDM physics and emphasize the unique discovery potential of power spectrum measurements in the high-redshift Universe.

\section*{Acknowledgements}
AR acknowledges discussions with Tal Adi. VG acknowledges the support from NASA through the Astrophysics Theory Program, Award Number 21-ATP21-0135, the support from the National Science Foundation (NSF) CAREER Grant No. PHY-2239205, and from the Research Corporation for Science Advancement under the Cottrell Scholar Program. The authors sincerely thank the referee for their insightful comments and suggestions that helped in improving the structure and quality of this paper.

\section*{Data Availability}
The data underlying this article will be shared on reasonable request to the authors.

\bibliographystyle{mnras}
\bibliography{bibliography}

\bsp
\label{lastpage}
\end{document}